\documentclass{emulateapj}

\usepackage{amsmath}
\usepackage{amssymb}
\usepackage{graphicx}
\usepackage{aas_macros}

\begin{document}

\title{Launching cosmic-ray-driven outflows from the magnetized interstellar medium}
\shorttitle{CR-driven Outflows}
\shortauthors{Girichidis et al.}

\author{Philipp~Girichidis\altaffilmark{1},
  Thorsten~Naab\altaffilmark{1},
  Stefanie~Walch\altaffilmark{2},
  Micha\l\ Hanasz\altaffilmark{3},
  Mordecai-Mark Mac Low\altaffilmark{4,5},
  Jeremiah~P.~Ostriker\altaffilmark{6},
  Andrea~Gatto\altaffilmark{1},
  Thomas~Peters\altaffilmark{1},
  Richard~W\"{u}nsch\altaffilmark{7},
  Simon~C.~O.~Glover\altaffilmark{5},
  Ralf~S.~Klessen\altaffilmark{5},
  Paul~C.~Clark\altaffilmark{8},
  Christian~Baczynski\altaffilmark{5},
}

\affil{$^1$Max-Planck-Institut f\"{u}r Astrophysik, Karl-Schwarzschild-Str. 1, 85741 Garching, Germany}
\affil{$^2$Physikalisches Institut, Universit\"{a}t zu K\"{o}ln, Z\"{u}lpicher Str. 77, 50937 K\"{o}ln, Germany}
\affil{$^3$Centre for Astronomy, Nicolaus Copernicus University, Faculty of Physics, Astronomy and Informatics, Grudziadzka 5, PL-87100 Toru\'n, Poland}
\affil{$^4$Department of Astrophysics, American Museum of Natural History, 79th Street at Central Park West, New York, NY 10024, USA}
\affil{$^5$Universit\"{a}t Heidelberg, Zentrum f\"{u}r Astronomie, Institut f\"{u}r Theoretische Astrophysik, Albert-Ueberle-Str. 2, 69120 Heidelberg, Germany}
\affil{$^6$Department of Astronomy, Columbia University, 1328 Pupin Hall, 550 West 120th Street, New York, New York 10027, USA}
\affil{$^7$Astronomical Institute, Academy of Sciences of the Czech Republic, Bocni II 1401, 141 31 Prague, Czech Republic}
\affil{$^8$School of Physics \& Astronomy, Cardiff University, 5 The Parade, Cardiff CF24 3AA, Wales, UK}

\begin{abstract}
We present a hydrodynamical simulation of the turbulent, magnetized, supernova (SN)-driven interstellar medium (ISM) in a stratified box that dynamically couples the injection and evolution of cosmic rays (CRs) and a self-consistent evolution of the chemical composition. CRs are treated as a relativistic fluid in the advection-diffusion approximation. The thermodynamic evolution of the gas is computed using a chemical network that follows the abundances of H$^+$, H, H$_2$, CO, C$^+$, and free electrons and includes (self-)shielding of the gas and dust. We find that CRs perceptibly thicken the disk with the heights of 90\% (70\%) enclosed mass reaching $\gtrsim1.5\,\mathrm{kpc}$ ($\gtrsim0.2\,\mathrm{kpc}$). The simulations indicate that CRs alone can launch and sustain strong outflows of atomic and ionized gas with mass loading factors of order unity, even in solar neighborhood conditions and with a CR energy injection per SN of $10^{50}\,\mathrm{erg}$, 10\% of the fiducial thermal energy of an SN. The CR-driven outflows have moderate launching velocities close to the midplane ($\lesssim100\,\mathrm{km\,s}^{-1}$) and are denser ($\rho\sim10^{-24}-10^{-26}\,\mathrm{g\,cm}^{-3}$), smoother, and colder than the (thermal) SN-driven winds. The simulations support the importance of CRs for setting the vertical structure of the disk as well as the driving of winds.
\end{abstract}

\keywords{cosmic rays --- ISM: structure --- ISM: jets and outflows --- magnetohydrodynamics (MHD) --- diffusion}

\section{Introduction}%
\label{sec:introduction}

Galactic outflows are important for the dynamical and chemical evolution of galaxies \citep[e.g.][]{VeilleuxCecilBlandHawthorn2005,OppenheimerDave2006}. They can enrich galactic halos with metals and regulate the baryonic angular momentum distribution in forming galaxies \citep[e.g.][]{UeblerEtAl2014}. Observations reveal that many actively star-forming galaxies in the early Universe $(z \gtrsim 0.5)$ show wind signatures \citep[e.g.][]{SteidelEtAl2010,NewmanEtAl2012,RubinEtAl2014}. In the local Universe, strong winds are mainly associated with starbursts (e.g. M82 or NGC 253). Milky Way (MW)-type galaxies show galactic fountain features. Despite their importance, the driving mechanisms of galactic winds are not precisely known. Galaxy-scale sub-grid modes including supernova (SN) explosions, radiation pressure, and stellar winds only partially succeed in explaining observed galactic wind properties.

Recent numerical simulations of the galactic interstellar medium (ISM;\citealt{SILCC1,GirichidisEtAl2015SILCC}) have shown that the properties of SN-driven outflows are tightly connected to details of the ISM structure and the environmental densities of SNe as one of the main driving mechanisms. Most previous calculations of the SN-driven ISM recover the net global quantities like mass fractions and volume filling factors of the different thermal states of the gas close to the disk midplane \citep{deAvillezBreitschwerdt2005,JoungMacLow2006,KimOstrikerKim2013,LiEtAl2015}. However, the vertical distribution of the gas above the disk as well as the temperature structure are not reproduced well by previous numerical models \citep{HenleyEtAl2010,WoodEtAl2010}.

Cosmic rays (CRs) as a non-thermal component (see, e.g. \citealt{Zweibel2013, GrenierBlackStrong2015}) might change the structures in the ISM even more. Their energy density is similar to the magnetic and thermal one (e.g., \citealt{Cox2005,Draine2011} and references therein). As CRs do not cool as efficiently as the thermal gas, they provide a long-lived energy reservoir. Their transport through the medium can be described by a diffusion process relative to the gas, which allows CRs to populate large regions of the disk creating a steady pressure gradient. Both facts make CRs a fundamentally different dynamical driving mechanism in the ISM. 

From theoretical considerations, a CR pressure gradient alone can generate outflows \citep{BreitschwerdtMcKenzieVoelk1993,EverettEtAl2008,DorfiBreitschwerdt2012}. This result is supported by recent galaxy-scale hydrodynamical simulations. \citet{HanaszEtAl2013} neglect the thermal impact of SNe and show that CRs alone can launch fast winds from the magnetized ISM in high surface density disks ($\Sigma_\mathrm{gas}=100\,M_\odot\mathrm{pc}^{-2}$). \citet{BoothEtAl2013} assume isotropic diffusion without the inclusion of magnetic fields and compare MW and Small Magellanic Cloud (SMC)-type galaxies, finding that CRs noticeably increase the mass loading factor ($\eta=\dot{M}/\mathrm{SFR}$) in SMC-like dwarf galaxies, but have a weaker impact in MW-type environments. With a similar implementation, \citet{SalemBryan2014} found stable outflows with $\eta=0.3$ but only for a high fraction of CR energy injection per SN. The main conclusion from these studies is that CRs can diffuse out of the dense regions of star formation and generate stable vertical pressure gradients. This helps to lift even dense and cold gas above the disk midplane. However, these models do not resolve the ISM structures and phases.

In this study, we present the first magnetohydrodynamical numerical simulations of stratified boxes that dynamically include CRs and follow the chemical evolution in the ISM. We aim for a more accurate description of the thermal and dynamical properties of a galactic disk with a focus on the vertical structure of the disk and the onset of outflows.

\section{Numerical Method and Simulation Setup}%
\label{sec:num-methods}

The simulations are performed with a modified version of the adaptive-mesh refinement code \textsc{FLASH} in version~4 \citep{FLASH00,DubeyEtAl2008}. We solve the magnetohydrodynamic (MHD) equations using the HLL3R method \citep{Waagan2009, Bouchut2010, Waagan2011}, extended to a separate monoenergetic CR fluid similar to \citet{YangEtAl2012}. Tests of this implementation are presented in \citet{GirichidisEtAl2015CR}.

The combined system of equations that we solve numerically is
\begin{align}
  \frac{\partial\rho}{\partial t} + \nabla\cdot\left(\rho\mathbf{v}\right) &= 0\\
  \frac{\partial\rho\mathbf{v}}{\partial t} + \nabla\cdot\left(\rho\mathbf{v}\mathbf{v}^\mathrm{T} - \frac{\mathbf{B}\mathbf{B}^\mathrm{T}}{4\pi}\right) + \nabla p_\mathrm{tot} &= \rho\mathbf{g}\\
  \frac{\partial e}{\partial t} + \nabla\cdot\left[\left(e + p_\mathrm{tot}\right)\mathbf{v} - \frac{\mathbf{B}(\mathbf{B}\cdot\mathbf{v})}{4\pi}\right] &= \notag\\
  \rho\mathbf{v}\cdot\mathbf{g} + \nabla\cdot\mathsf{K}\nabla e_{_\mathrm{CR}} + \dot{u}_\mathrm{chem} + \dot{u}_\mathrm{inj}&\\
  \frac{\partial\mathbf{B}}{\partial t} - \nabla \times \left(\mathbf{v}\times\mathbf{B}\right) &= 0\\
  \frac{\partial e_{_\mathrm{CR}}}{\partial t} + \nabla\cdot(e_{_\mathrm{CR}}\mathbf{v}) &= \notag\\
  -p_{_\mathrm{CR}}\nabla\cdot\mathbf{v} + \nabla\cdot(\mathsf{K}\nabla e_{_\mathrm{CR}}) + Q_{_\mathrm{CR}},
\end{align}

Here, $\rho$ is the mass density, $\mathbf{v}$ is the velocity, and $\mathbf{B}$ is the magnetic field. The total energy density, 
\begin{equation}
    e = \rho v^2/2 + e_\mathrm{th} + e_{_\mathrm{CR}} + B^2/8\pi,
\end{equation}
includes kinetic, thermal, CR, and magnetic contributions. We evolve the CR energy density, $e_{_\mathrm{CR}}$, separately. The total pressure is
\begin{align}\label{eq:total-pressure}
  p_\mathrm{tot} &= p_\mathrm{th} & + &\,p_{_\mathrm{CR}} & + &\,p_\mathrm{mag}\\
  &= (\gamma-1)e_\mathrm{th} & + &\,(\gamma_{_\mathrm{CR}}-1)e_{_\mathrm{CR}} & + &\,B^2/8\pi.
\end{align}
The closure relation for the system, the equation of state, combines the different contributions from CR and thermal pressure in an effective adiabatic index, $\gamma_\mathrm{eff}$, 
\begin{equation}
  \gamma_\mathrm{eff} = \frac{\gamma p_\mathrm{th} + \gamma_{_\mathrm{CR}} p_{_\mathrm{CR}}}{p_\mathrm{th} + p_{_\mathrm{CR}}},
\end{equation}
where we set $\gamma = 5/3$ and $\gamma_\mathrm{CR} = 4/3$ for gas and CRs, respectively. For the CR diffusion tensor, $\mathsf{K}$, we assume a value of $10^{28}\,\mathrm{cm^2\,s^{-1}}$ parallel and $10^{26}\,\mathrm{cm^2\,s^{-1}}$ perpendicular to the magnetic field lines \citep[e.g.,][]{StrongEtAl2007,NavaGabici2013}.

An exact treatment of CR propagation and the coupling to the gas would include the generation of Alfv\'{e}n waves due to CR streaming and wave damping due to ion-neutral collisions and nonlinear Landau damping including anisotropic particle distribution functions \citep[e.g.][]{KulsrudPearce1969,PtuskinEtAl1997,BreitschwerdtEtAl2002,DorfiBreitschwerdt2012,UhligEtAl2012,WienerZweibelOh2013}. However, the detailed coupling of CRs to the gas -- in particular, the weakly ionized gas -- is an open question and beyond the scope of this Letter. For our model we assume that CRs are scattered sufficiently by small-scale background turbulence, which we do not resolve. We therefore adopt the simplified transport equation following \citet{SchlickeiserLerche1985}, see also \citet{YangEtAl2012,HanaszEtAl2013}, which is derived from an isotropic particle distribution function.

Radiative and chemical heating and cooling, $\dot{u}_\mathrm{chem}$, are computed using a chemical network that follows the abundances of H$^{+}$, H, H$_{2}$, C$^{+}$, CO, and free electrons \citep[see][and references therein]{GloverClark2012b}. We assume a constant CR ionization rate ($3\times10^{-17}\,\mathrm{s}$). We follow (self-)shielding of the molecular gas as described in \citet{ClarkGloverKlessen2012}. Photoionization is neglected. The thermal and CR energy injected by the SNe are included in $\dot{u}_\mathrm{inj}$ and $Q_\mathrm{CR}$. The external gravitational acceleration, $\mathbf{g}$, is taken from \citet{KuijkenGilmore1989}.

The stratified disk is initially atomic, at rest, and in pressure equilibrium with a scale height of $60\,\mathrm{pc}$ and a gas surface density of $10\,M_\odot\mathrm{pc}^{-2}$ (see \citealt{SILCC1} for further details) in a box of $2\times2\times\pm20\,\mathrm{kpc}$. For $|z|<2.5\,\mathrm{kpc}$, we adopt a cell size of $15.6\,\mathrm{pc}$, which corresponds to an effective numerical resolution of $128\times128\times2560$ cells. For $|z|>2.5\,\mathrm{kpc}$, the adaptive-mesh refinement can coarsen the resolution. Motivated by the Kennicutt-Schmidt (KS) relation \citep{Kennicutt1998}, we explode SNe at a fixed rate of $60\,\mathrm{Myr}^{-1}\mathrm{kpc}^{-2}$ assuming one massive star per $100\,M_\odot$. Similar to \citet{deAvillezBreitschwerdt2004} or \citet{JoungMacLow2006}, we assign 20\% of the explosions to SN type~Ia and 80\% to SN type~II with scale heights of $325\,\mathrm{pc}$ and $50\,\mathrm{pc}$, respectively \citep{TammannLoefflerSchroeder1994}. However, estimates of the displacement of runaway OB stars suggests that the effective scale height of type~II SNe might be larger \citep{LiEtAl2015}.

We separately vary the thermal energy injection per SN, $E_\mathrm{th}=f_\mathrm{th}E_\mathrm{SN}$, and the CR energy injection, $E_\mathrm{CR}=f_\mathrm{CR}E_\mathrm{SN}$, with $E_\mathrm{SN}=10^{51}\,\mathrm{erg}$. We discuss three different simulations: purely thermal SNe ($f_\mathrm{th}=1.0, f_\mathrm{CR}=0.0$, \texttt{TH1-CR0}), purely CR SNe ($f_\mathrm{th}=0.0, f_\mathrm{CR}=0.1$, \texttt{TH0-CR1}), and the physically most realistic setup with both energy injections ($f_\mathrm{th}=1.0, f_\mathrm{CR}=0.1$, \texttt{TH1-CR1}).

\section{Morphological Evolution}%
\label{sec:evolution}

\begin{figure*}
  \begin{minipage}{\textwidth}
    \centering
    \includegraphics[width=\textwidth]{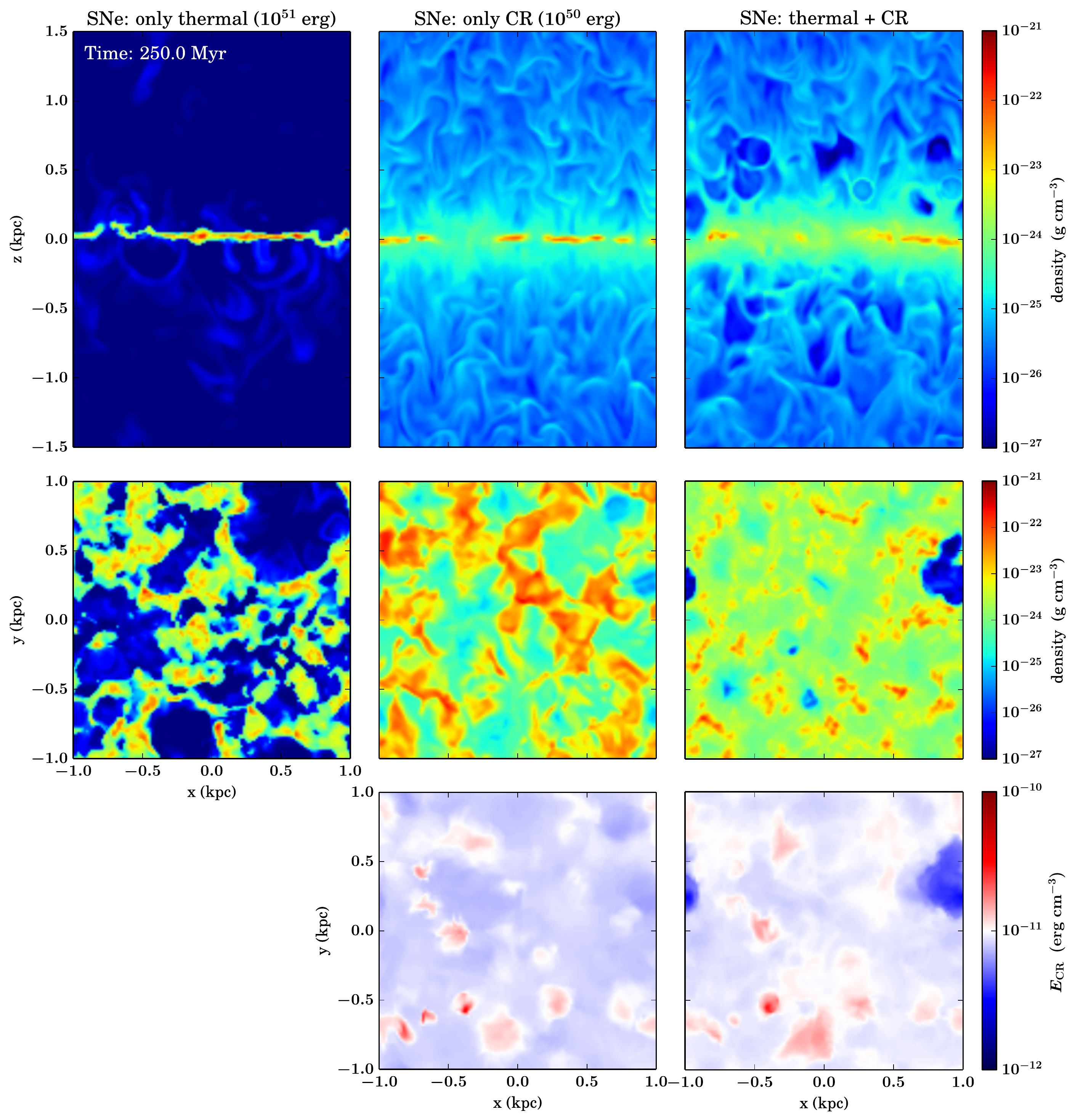}
    \caption{Structure of the disk after $250\,\mathrm{Myr}$ of evolution. The left panel shows the run without CRs, the middle panel the run with purely CR feedback, and the right panel the simulation with both feedback energies. In the top and middle row, we plot the density edge-on and face-on in cuts through the center of the box. The bottom row depicts the CR energy density in the midplane. The disk in the run including CRs is significantly thicker compared to the purely thermal run. The thermal energy input of the SNe is primarily important in the denser part of the disk. CRs show their main impact in generating an extended atmosphere. The CR energy density is smooth in the midplane. Local variations are due to recent SNe and the resulting CR injection. Those variations disappear after less than a Myr due to diffusion.}
    \label{fig:sim-overview}
  \end{minipage}
\end{figure*}

In all simulations, the SNe induce the formation of filaments, clumps, and voids. The presence of CRs leads to a smoother morphological appearance. An overview of the disk structure after $250\,\mathrm{Myr}$ is depicted in Fig.~\ref{fig:sim-overview} with cuts through the center of the box for the density edge-on (top) and face-on (middle) and the CR energy density (bottom).
From left to right, we present \texttt{TH1-CR0}, \texttt{TH0-CR1}, and \texttt{TH1-CR1}. Both runs with CRs quickly push gas away from the midplane and form an extended atmosphere, whereas in the simulation with only thermal SN feedback, the disk remains compact without a noticeable extended layer of gas above the disk. The CR energy density shows relatively variations of about one order of magnitude. High CR energy peaks correspond to recent SNe and disappear quickly due to diffusion.

\begin{figure}
  \includegraphics[width=8cm]{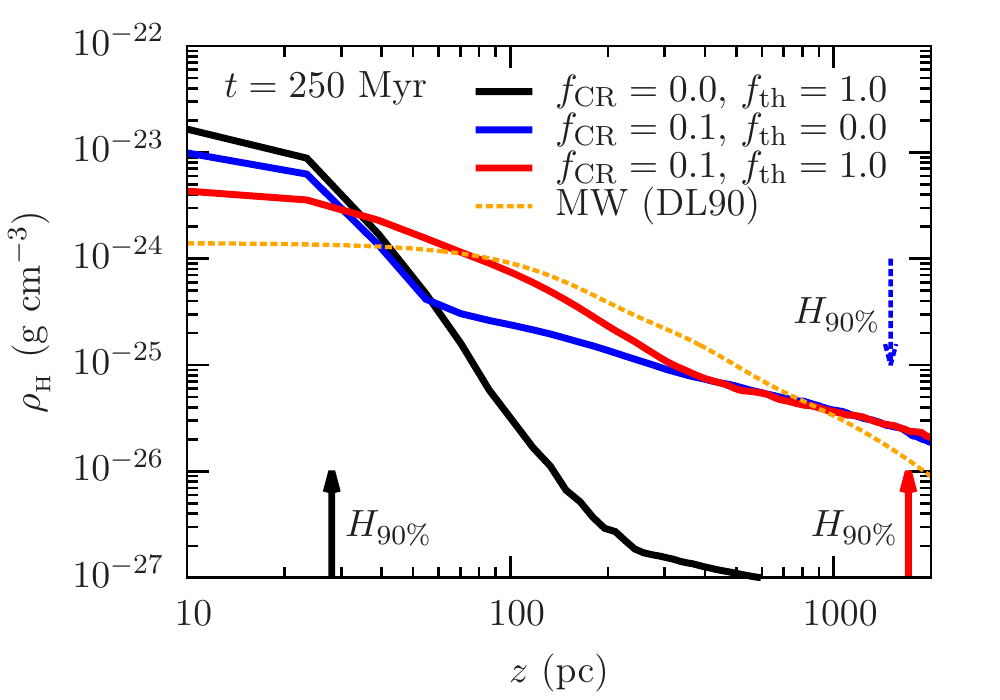}
  \caption{Vertical profiles of the total gas density for all simulations. The arrows indicate the height of 90\% enclosed mass. A fit to the observed density profile of the solar neighborhood \citep{DickeyLockman1990} are shown in yellow. Thermal energy injection alone leads to a compact atomic gas distribution. Including CR feedback results in very extended distributions, which are much closer to the observed extent of the gas. The profiles indicate that CRs have their main impact at larger altitudes.}
  \label{fig:dens-profiles}
\end{figure}

\begin{figure*}
  \begin{minipage}{\textwidth}
    \centering
    \includegraphics[width=\textwidth]{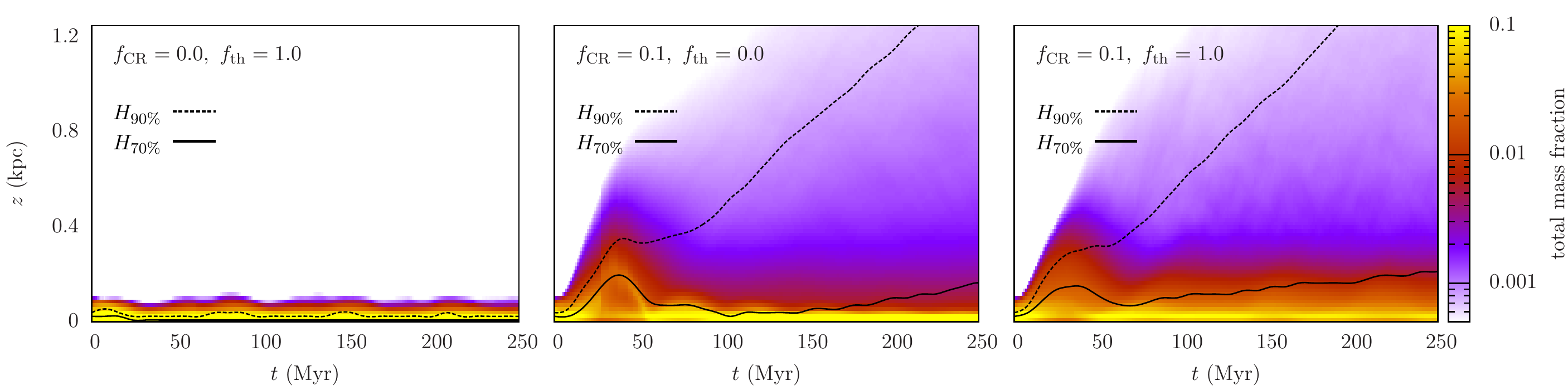}
    \caption{Time evolution of the vertical gas distribution. Color coded is the total gas fraction. Overplotted are the heights of 90\% ($H_{90\%}$, dashed line) and $70\%$ ($H_{70\%}$, solid line) enclosed total mass. In the purely thermal run (left) the disk does not show variations over time. Including CRs leads to a slow expansion of bulk of the disk ($H_{70\%}$) and a fast expansion of the envelope ($H_{90\%}$).}
    \label{fig:HI-fraction-time-evol}
  \end{minipage}
\end{figure*}

In Fig.~\ref{fig:dens-profiles}, we show the vertical profile of the density for all simulations at $t=250\,\mathrm{Myr}$. The arrows indicate the heights of 90\% enclosed mass. The inclusion of CRs increases $H_{90\%}$ from $\sim30\,\mathrm{pc}$ to $\sim1500\,\mathrm{pc}$ with similar values for both runs including CRs. The additional thermal energy injection from SNe in run \texttt{TH1-CR1} changes the profile up to a height of $\sim300\,\mathrm{pc}$. At larger altitudes, the profiles for \texttt{TH0-CR1} and \texttt{TH1-CR1} are basically indistinguishable. For comparison, we include the vertical density profile of the MW \citep{DickeyLockman1990} showing reasonable agreement with both CR runs.

The time evolution of the disk thickness is shown in Fig.~\ref{fig:HI-fraction-time-evol} including $70\%$ (solid line) and $90\%$ (dashed line) of enclosed mass. Run \texttt{TH1-CR0} shows little variation of the disk over the entire simulation time. For the other runs, the CR energy density can push a significant fraction of the gas to heights of up to $\sim400\,\mathrm{pc}$ in the beginning of the simulation before the initially ordered magnetic field becomes tangled (Parker instability) and the CR energy quickly diffuses along the opened up field lines. Opening magnetic field lines enable the reduction of CR pressure via the escape of excessive CR gas initially trapped in the horizontal magnetic field. The bulk of the disk thickens slowly over time ($H_{70\%}$), whereas the envelope expands quickly ($H_{90\%}$).

\section{Outflows}

\begin{figure}
  \includegraphics[width=8cm]{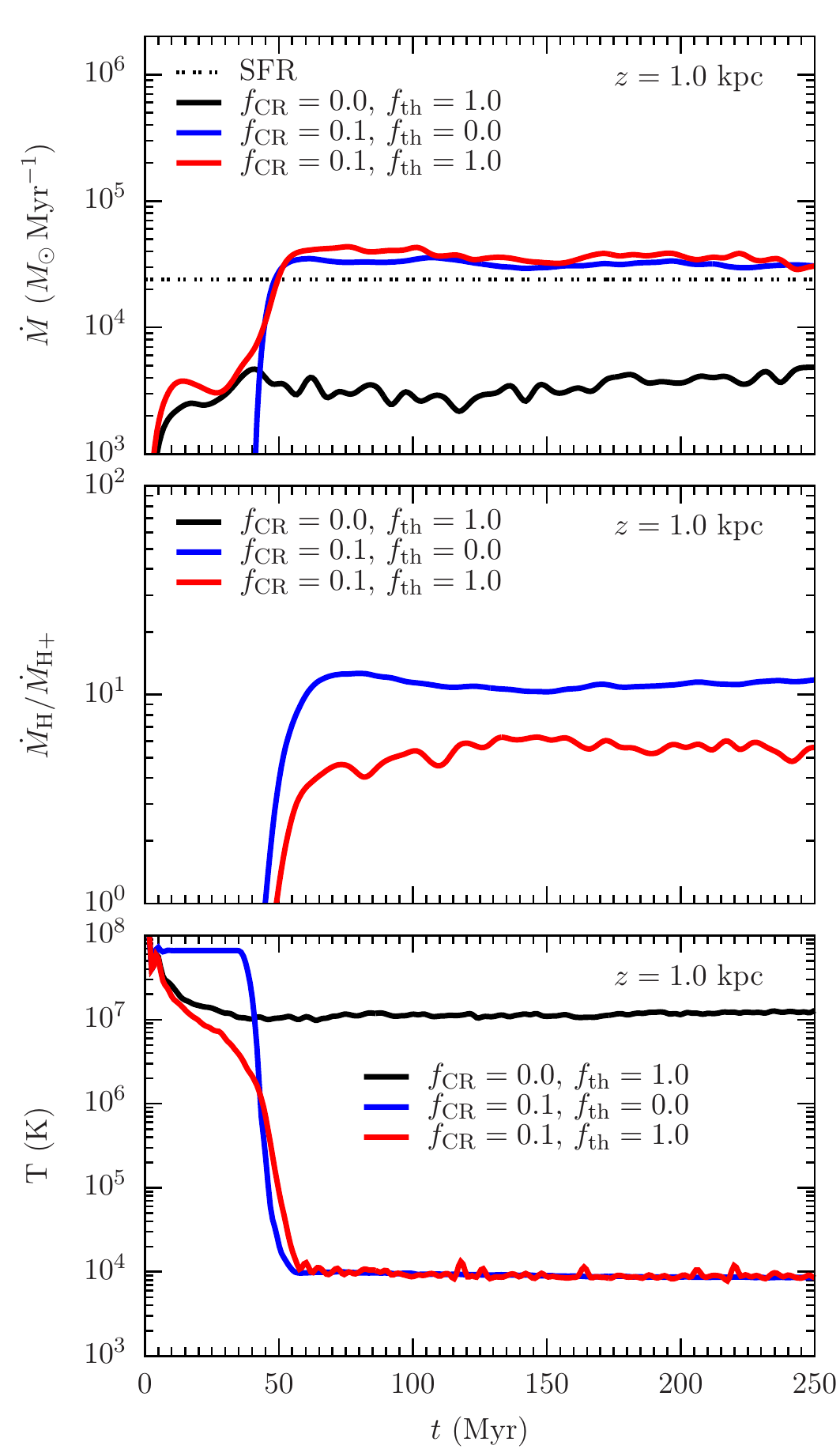}
  \caption{Outflow rate (top), the ratio of atomic to ionized hydrogen (middle), and the temperature (bottom) at $z=\pm1\,\mathrm{kpc}$. All simulations show outflows. The purely thermal SNe have outflow rates lower than the star formation rate. Including CRs, the mass loading factor increases to unity. The outflows driven by purely thermal SNe are hot ($\sim10^7\,\mathrm{K}$) and consist of ionized hydrogen. The CR-driven counterparts are colder ($10^4\,\mathrm{K}$) and mainly composed of atomic hydrogen.}
  \label{fig:outflows}
\end{figure}

The properties of the outflows are plotted in Fig.~\ref{fig:outflows}, with the total outflow rate (top), the ratio of atomic to ionized hydrogen (middle), and the temperature (bottom), averaged at $\pm1\,\mathrm{kpc}$. Included is all the gas with velocities pointing away from the midplane. All simulations drive outflows at a roughly constant rate after $\sim 50\,\mathrm{Myr}$. Run \texttt{TH1-CR0} reaches a mass loading factor, $\eta=\dot{M}/\mathrm{SFR}\sim0.1-0.2$ at $1\,\mathrm{kpc}$ above the midplane. Including CRs increases $\eta$ to order unity. This indicates that in our setups the main driver for outflows is CRs rather than the thermal contribution from the SNe. The chemical composition, however, is affected by the thermal contribution. The outflowing gas in the simulation with purely thermal SN feedback is very hot ($10^7\,\mathrm{K}$; bottom panel of Fig.~\ref{fig:outflows}) and composed entirely of ionized hydrogen H$^+$ (not visible). In the other two cases with CRs, the outflowing gas is warm ($\sim10^4\,\mathrm{K}$) and mainly composed of atomic hydrogen with $\dot{M}_\mathrm{H}/\dot{M}_\mathrm{H+}\sim5-10$. The inclusion of photoionization feedback is likely to decrease this value, but will not significantly alter the temperature of the outflowing gas or the very low volume-filling fractions of the hot gas. As soon as the CR pressure gradient has built up, significant fractions of neutral and warm gas can be accelerated away from the disk midplane \citep[see, e.g.][]{SalemBryan2014}.

\begin{figure}
  \centering
  \includegraphics[width=8cm]{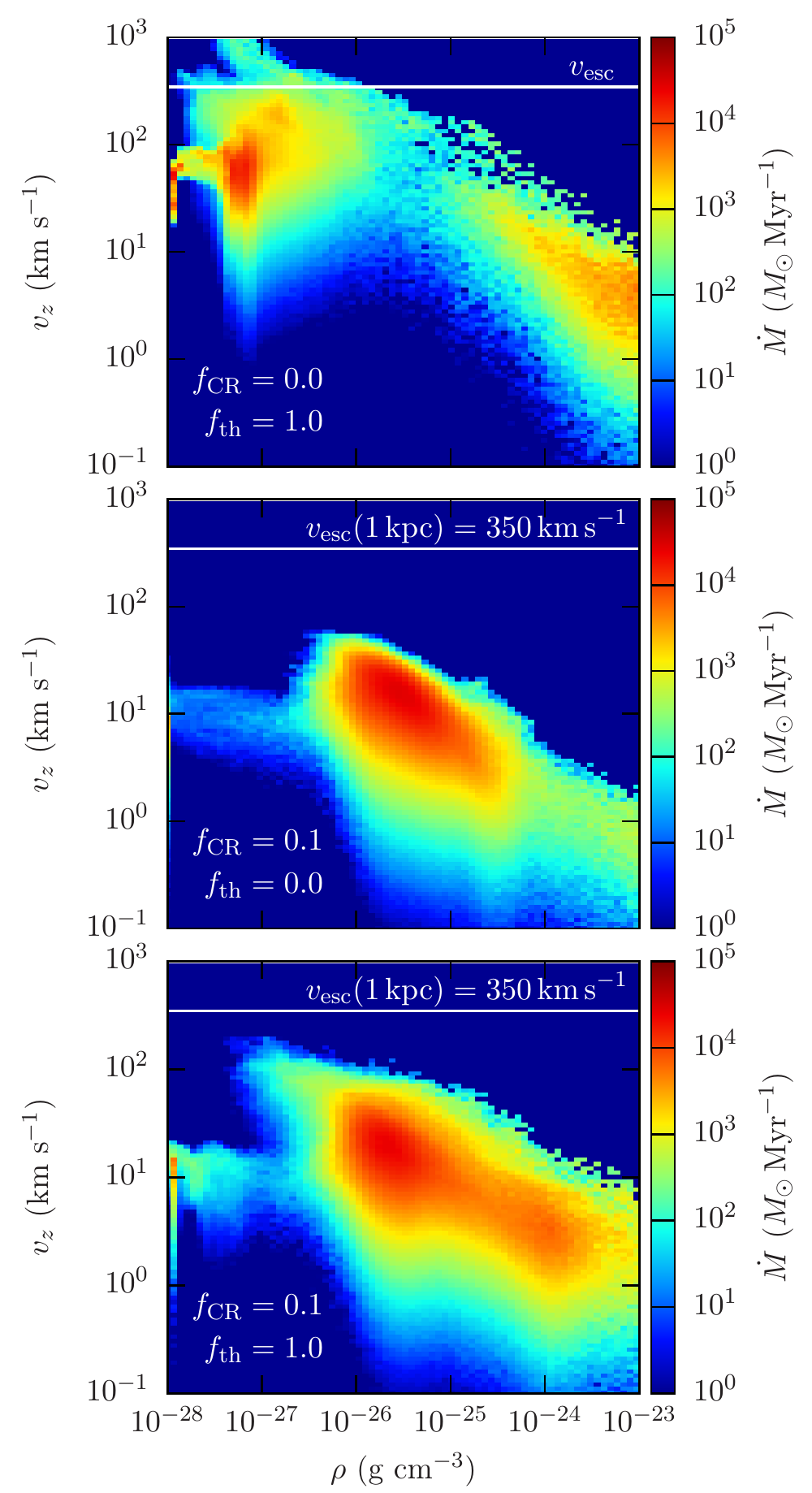}
  \caption{Histograms of the outflows ($v_zz>0$) as a function of density and velocity. Purely thermal SN feedback (top) leads to fast low-density outflows. Including CRs (middle and bottom) the transported gas is two orders of magnitude denser and a factor of a few slower with broader distributions}
  \label{fig:outflow-phaseplots}
\end{figure}

The connection between density and outflow velocity is depicted in Fig.~\ref{fig:outflow-phaseplots}. For \texttt{TH1-CR0} (top panel), the SNe explode partially in regions of very low density, can thus expand to large radii, and push gas out of the midplane at very high velocities of up to $\sim500\,\mathrm{km\,s}^{-1}$ at very low densities below $\rho\sim10^{-26}\,\mathrm{g\,cm}^{-3}$. If CRs are included, gas is slowly transported away from the dense regions in the midplane to form an extended atmosphere, whose flow of gas becomes perceptibly denser and overall slower compared to \texttt{TH1-CR0}. At the low heights investigated here, the gas does not reach escape velocity. Again, this is because the CRs quickly diffuse throughout the volume and build up a stable, slowly varying energy (pressure) gradient along the $z$-direction that slowly but continuously accelerates the gas, in contrast to the short-lived locally acting thermal energy input. This underlines the fundamental difference of galactic winds driven by SNe and CRs. The comparison between \texttt{TH0-CR1} (middle) and the combined energy input (\texttt{TH1-CR1}) shows only little impact on the bulk density of the outflow ($\rho\sim10^{-26}-10^{-24}\,\mathrm{g\,cm}^{-3}$). The contribution of the thermal component in the SN driving increases the peak velocities slightly from $\sim50\,\mathrm{km\,s}^{-1}$ to $\sim100\,\mathrm{km\,s}^{-1}$. However, the dense atmosphere around the midplane, in which the SNe explode, reduces their net impact on the velocities compared to \texttt{TH1-CR0} \citep[see also][]{GirichidisEtAl2015SILCC}. For observational signatures in the soft X-ray emission, see \citet{PetersEtAl2015}.

\section{Discussion}

We find that CRs are able to perceptibly increase the thickness of the disk and launch and sustain strong outflows with mass loading factors of order unity. For the simulations presented here, the CR-driven winds are slower ($50-100\,\mathrm{km\,s^{-1}}$) and denser than the (pure thermally) SN-driven winds. This is qualitatively consistent with the results by \citet{BoothEtAl2013}. \citet{SalemBryan2014} report mass loading factors below unity for their fiducial model with a total CR fraction per SN of 30\% rather than our 10\%. The simulations of \citet{HanaszEtAl2013} also show mass loading factors above unity but for massive galactic disks with high SN rates. We note, however, that an important parameter of the model is the exact value of the assumed CR diffusion coefficient. Its value controls both the terminal wind speed and mass-loss rate \citep{DorfiBreitschwerdt2012}. For small values of $\mathsf{K}$, mass-loss rates are higher at lower speeds, whereas large diffusion coefficients result in lower mass loading factors but higher velocities. We found a similar behavior in tests with varying diffusion tensor for the setup described here. Full details will be provided in a follow-up paper.

Previous models of stratified disks have shown that magnetic fields help to increase the thickness of the disk \citep{HillEtAl2012}, but that they alone are not strong enough to reach observed scale heights. The combined MHD-CR simulation presented here seems a promising solution to this problem. 

We also note that the impact of purely thermal SNe strongly depends on the positioning of the SNe and therefore might be resolution dependent. In particular, the SNe that explode in density peaks have a weaker net impact \citep{GattoEtAl2015,SILCC1,GirichidisEtAl2015SILCC}, and thus the outflow rates in the purely thermal run, \texttt{TH1-CR0}, might be lower limits. This weak SN efficiency might also be responsible for small fractions of hot gas in the CR runs. However, we find that the inclusion of CRs results in outflows that are compatible with the observations. This suggests that CRs will always play an important role in the matter cycle of galaxies.

\section{Conclusions}
We present the first ISM simulations that dynamically couple CRs with the chemodynamical evolution of the magnetized ISM. The CRs are included in an advection-diffusion approach with an anisotropic diffusion tensor. We follow the formation of density structures, the vertical distribution of gas in a galactic disk, and the launching of outflows in three different runs varying the SN feedback between purely thermal, purely CR, and combined thermal and CR feedback. Our conclusions can be summarized as follows:
\begin{enumerate}
  \item Including CRs thickens the galactic disk. The height of $90\%$ enclosed total mass is found to be $\sim1.5\,\mathrm{kpc}$ in the case of 10\% CR energy injection per SN after $250\,\mathrm{Myr}$ and to increase continuously. Comparison with the vertical density distribution in the MW indicates good agreement.
  \item We find that CRs quickly lead to the formation of a warm and neutral galactic atmosphere providing a mass reservoir for galactic winds and outflows. Whereas the thermal contribution of the SNe mainly shapes the disk close to the midplane, the additional CR energy shows the strongest impact above the disk and in the halo.
  \item All simulations drive gas out of the midplane with little variation over time.
    For purely thermal SN feedback, the outflows are hot and composed of mainly ionized hydrogen with rates below the star formation rate. They are fast (up to $\sim\mathrm{~a~few~}100\,\mathrm{km\,s}^{-1}$) with low densities ($\rho\lesssim10^{-27}\,\mathrm{g\,cm}^{-3}$). CRs alone can drive outflows with mass loading factors of order unity, which are warm ($10^4\,\mathrm{K}$) and mainly composed of atomic hydrogen. They are a factor of a few slower ($\sim10-50\,\mathrm{km\,s}^{-1}$) and $1-2$ orders of magnitude denser ($\rho\sim10^{-26}-10^{-25}\,\mathrm{g\,cm}^{-3}$) compare to their thermally driven counterparts.
\end{enumerate}

\section*{Acknowledgements}
We thank the referee for valuable comments and questions.
P.G., S.W., T.N., T.P., A.G., S.C.O.G., R.S.K., and C.B. acknowledge support from the DFG Priority Program 1573 {\em Physics of the Interstellar Medium}.
S.W. acknowledges the support of the Bonn-Cologne Graduate School, which is funded through the Excellence Initiative.
T.N. acknowledges support from the DFG cluster of excellence \emph{Origin and Structure of the Universe}.
R.W. acknowledges support by the Czech Science Foundation grant 209/12/1795 and by the project RVO:67985815 of the Academy of Sciences of the Czech Republic.
R.S.K., S.C.O.G., and C.B. thank the DFG for funding via the SFB 881 The Milky Way System (subprojects B1, B2, and B8). R.S.K. furthermore acknowledges support from the European Research Council under the European Community’s Seventh Framework Programme (FP7/2007-2013) via the ERC Advanced Grant STARLIGHT (project number 339177).
M.-M.M.L. was partly supported by NSF grant AST11-09395 and the Alexander von Humboldt Foundation.
M.H. acknowledges support from Polish NCN grant N203 511038.
The authors thank the Max Planck Computing and Data Facility (MPCDF) for computing time and data storage.
The software used in this work was developed in part by the DOE NNSA ASC- and DOE Office of Science ASCR-supported Flash Center for Computational Science at the University of Chicago. 

\bibliographystyle{mn2e}
\bibliography{astro.bib}
\end{document}